\documentstyle[charm2000]{article}
\def \arnps#1#2#3{{\it Ann. Rev. Nucl. Part. Sci.} {\bf#1} (#3) #2}

\def \ba88{{\it Particles and Fields 3} (Proceedings of the 1988 Banff Summer
Institute on Particles and Fields), edited by A. N. Kamal and F. C. Khanna
(World Scientific, Singapore, 1989)}

\def \be87{{\it Proceedings of the Workshop on High Sensitivity Beauty
Physics at Fermilab,} Fermilab, Nov. 11-14, 1987, edited by A. J. Slaughter,
N. Lockyer, and M. Schmidt (Fermilab, Batavia, IL, 1988)}
\def \cn{Collaboration}
\def \corn{{\it Proceedings of the XVI International Symposium on Lepton and
Photon Interactions}, Cornell University, August 10--15, 1993, edited by P.
Drell and D. Rubin (New York, AIP, 1994)}
\def \cp89{{\it CP Violation,} edited by C. Jarlskog (World Scientific,
Singapore, 1989)}
\def \dpf91{{\it The Vancouver Meeting - Particles and Fields '91}
(Division of Particles and Fields Meeting, American Physical Society,
Vancouver, Canada, Aug.~18-22, 1991), ed. by D. Axen, D. Bryman, and M. Comyn
(World Scientific, Singapore, 1992)}

\def \efi{Enrico Fermi Institute Report No.~}
\def \hb87{{\it Proceeding of the 1987 International Symposium on Lepton and
Photon Interactions at High Energies,} Hamburg, 1987, ed. by W. Bartel
and R. R\"uckl (Nucl.~Phys.~B, Proc. Suppl., vol. 3) (North-Holland,
Amsterdam, 1988)}

\def \ibj#1#2#3{{\it ibid.} {\bf#1} (#3) #2}

\def \ite{{\it et al.}}

\def \ky85{{\it Proceedings of the International Symposium on Lepton and
Photon Interactions at High Energy,} Kyoto, Aug.~19-24, 1985, edited by M.
Konuma and K. Takahashi (Kyoto Univ., Kyoto, 1985)}
\def \lat90{{\it Results and Perspectives in Particle Physics} (Proceedings of
Les Rencontres de Physique de la Vallee d'Aoste [4th], La Thuile, Italy, Mar.
18-24, 1990), edited by M. Greco (Editions Fronti\`eres, Gif-Sur-Yvette,
France,
1991)}
\def \lg91{International Symposium on Lepton and Photon Interactions, Geneva,
Switzerland, July, 1991}
\def \lkl87{{\it Selected Topics in Electroweak Interactions} (Proceedings of
the Second Lake Louise Institute on New Frontiers in Particle Physics, 15 --
21 February, 1987), edited by J. M. Cameron \ite~(World Scientific, Singapore,
1987)}
\def \mpla #1#2#3{{\it Mod. Phys. Lett.} A {\bf#1} (#3) #2}

\def \np#1#2#3{{\it Nucl. Phys.} {\bf#1} (#3) #2}
\def \oxf65{{\it Proceedings of the Oxford International Conference on
Elementary Particles} 19/25 Sept.~1965, ed.~by T. R. Walsh (Chilton, Rutherford
High Energy Laboratory, 1966)}

\def \pl#1#2#3{{\it Phys. Lett.} {\bf#1} (#3) #2}
\def \plb#1#2#3{{\it Phys. Lett.} B {\bf#1} (#3) #2}

\def \prd#1#2#3{{\it Phys. Rev.} D {\bf#1} (#3) #2}
\def \prl#1#2#3{{\it Phys. Rev. Lett.} {\bf#1} (#3) #2}

\def \ptp#1#2#3{{\it Prog. Theor. Phys.} {\bf#1} (#3) #2}
\def \rmp#1#2#3{{\it Rev. Mod. Phys.} {\bf#1} (#3) #2}

\def \si90{25th International Conference on High Energy Physics, Singapore,
Aug. 2-8, 1990, Proceedings edited by K. K. Phua and Y. Yamaguchi (World
Scientific, Teaneck, N. J., 1991)}
\def \slac75{{\it Proceedings of the 1975 International Symposium on
Lepton and Photon Interactions at High Energies,} Stanford University, Aug.
21-27, 1975, edited by W. T. Kirk (SLAC, Stanford, CA, 1975)}
\def \slc87{{\it Proceedings of the Salt Lake City Meeting} (Division of
Particles and Fields, American Physical Society, Salt Lake City, Utah, 1987),
ed. by C. DeTar and J. S. Ball (World Scientific, Singapore, 1987)}
\def \smass82{{\it Proceedings of the 1982 DPF Summer Study on Elementary
Particle Physics and Future Facilities}, Snowmass, Colorado, edited by R.
Donaldson, R. Gustafson, and F. Paige (World Scientific, Singapore, 1982)}
\def \smass90{{\it Research Directions for the Decade} (Proceedings of the
1990 DPF Snowmass Workshop), edited by E. L. Berger (World Scientific,
Singapore, 1991)}
\def \smassb{{\it Proceedings of the Workshop on $B$ Physics at Hadron
Accelerators}, Snowmass, Colorado, 21 June -- 2 July 1994, ed.~by P. McBride
and C. S. Mishra, Fermilab report FERMILAB-CONF-93/267 (Fermilab, Batavia, IL,
1993)}
\def \stone{{\it B Decays}, edited by S. Stone (World Scientific, Singapore,
1994)}
\def \tasi90{{\it Testing the Standard Model} (Proceedings of the 1990
Theoretical Advanced Study Institute in Elementary Particle Physics),
edited by M. Cveti\v{c} and P. Langacker (World Scientific, Singapore, 1991)}

\def \zhetf#1#2#3#4#5#6{{\it Zh. Eksp. Teor. Fiz.} {\bf #1} (#3) #2 [Sov.
Phys. - JETP {\bf #4} (#6) #5]}

\def \zpc#1#2#3{{\it Zeit. Phys.} C {\bf#1} (#3) #2}

\begin{document}
\title{What Charm Can Tell Us About Beauty}

\author{Jonathan L. Rosner		% **** 1st line of author names
 \\ {\sl Enrico Fermi Institute and Department of Physics}
 \\ {\sl University of Chicago, Chicago, IL 60637}
         }  % end of \author
% Note: only 3 inches TOTAL allowed for title and authors!!
%
\maketitle

\vspace{-3.6in}
\rightline{EFI 94-33}
\rightline{hep-ph/9407256}
\rightline{July 1994}
\vspace{-0.6in}
\leftline{Presented at CHARM2000 Workshop}
\leftline{Fermilab, June 7 -- 9, 1994}
\vspace{2.1in}

\begin{abstract}
A number of ways are reviewed in which the study of charmed particles can
answer corresponding questions about particles containing $b$ quarks.  Topics
include the properties of resonances, the magnitude of decay constants, the
size of spin-dependent effects, and the hierarchy of lifetime differences.
\end{abstract}

\section{Introduction}

The study of charmed particles is of interest not only in its own right, but
for the information it can provide about particles containing $b$ quarks.

Charmed particles are relatively easy to produce.  In the standard electroweak
picture, their weak decays are unlikely to exhibit detectable CP-violating
effects, and are noticeably affected by strong interactions.  The good news is
that these strong interactions are rich and easily studied.

Particles containing $b$ quarks are much harder to produce.  Their weak
interactions (again, in the conventional view) are expected to be a rich source
of observable CP-violating phenomena, and to be less polluted by the strong
interactions.  However, these strong interactions are still important (for
example, one needs to know $B$ meson decay constants), but their study is
hampered by a lack of statistics.  Here, charmed particles can be very helpful.

Many questions regarding $B$ hadrons can benefit from the corresponding studies
of charmed particles.  These include resonances, spin-dependent effects,
lifetime differences, and form factors for heavy-to-light weak transitions.
Moreover, since weak decays of $B$ hadrons often involve charm, the branching
ratios of charmed particles are crucial in determining the corresponding
$B$ branching ratios.

This brief article touches upon some of the ways in which information about
charmed particles can be applied to the corresponding states containing $b$
quarks.  In Section 2 we review the relevant aspects of heavy quark symmetry
permitting an extrapolation from charm to beauty.  Section 3 is devoted to
the open questions facing the study of CP violation in $B$ decays, with
emphasis on parallels with charm.  Section 4 is devoted to strange $B$'s:
their production, masses, and mixings, and the corresponding questions for
charm.  Heavy meson decay constants, for which we have partial information
in the case of charm, are treated in Section 5.  Heavy baryon spectra are
discussed in Section 6, while Section 7 treats lifetime differences.  We
summarize in Section 8.

\section{Heavy quark symmetry}

In a hadron containing a single heavy quark, that quark ($Q = c$ or $b$) plays
the role of an atomic nucleus, with the light degrees of freedom (quarks,
antiquarks, gluons) analogous to the electron cloud.  The properties of hadrons
containing $b$ quarks (we shall call them $B$ hadrons) then can calculated from
the corresponding properties of charmed particles by taking account \cite{HQS}
of a few simple ``isotope effects.''  For example, if $q$ denotes a light
antiquark, the mass of a $Q \bar q$ meson can be expressed as
\begin{equation}
M(Q \bar q) = m_Q + {\rm const.}[n,\ell] + \frac{\langle p^2 \rangle}{2 m_Q} +
a \frac{\langle {\bf \sigma_q \cdot \sigma_Q} \rangle}{m_q m_Q} + {\cal O}
(m_Q^{-2})~~~.
\end{equation}
Here the constant depends only on the radial and orbital quantum numbers $n$
and $\ell$.  The $\langle p^2 \rangle /2m_Q$ term expresses the dependence of
the heavy quark's kinetic energy on $m_Q$, while the last term is a hyperfine
interaction.  The expectation value of $\langle {\bf \sigma_q \cdot \sigma_Q}
\rangle$ is $(+1,~-3)$ for $J^P = (1^-,~0^-)$ mesons. If we define
$\overline{M} \equiv [3 M(1^-) + M(0^-)]/4$, we find
\begin{equation}
m_b - m_c + \frac{\langle p^2 \rangle}{2 m_b} - \frac{\langle p^2 \rangle}
{2 m_c} = \overline{M}(B \bar q) - \overline{M}(c \bar q) \simeq 3.34~{\rm
GeV}~~.
\end{equation}
so $m_b - m_c > 3.34$ GeV, since $\langle p^2 \rangle > 0$.  Details of this
picture which are of interest include (1) the effects of replacing nonstrange
quarks with strange ones, (2) the energies associated with orbital excitations,
(3) the size of the $\langle p^2 \rangle$ term, and (4) the magnitude of
hyperfine effects.  In all cases there exist ways of using information about
charmed hadrons to predict the properties of the corresponding $B$ hadrons.

\section{CP violation and $B$ mesons}

\subsection{The CKM matrix}

\subsubsection{Parameters and their values}

In a parametrization \cite{wp} in which the rows of the CKM \cite{cab,KM}
matrix are labelled by $u,~c,~t$ and the columns by $d,~s,~b$, we may write
\begin{equation}
V = \left ( \begin{array}{c c c}
V_{ud} & V_{us} & V_{ub} \\
V_{cd} & V_{cs} & V_{cb} \\
V_{td} & V_{ts} & V_{tb}
\end{array} \right )
\approx \left [ \matrix{1 - \lambda^2 /2 & \lambda & A \lambda^3 ( \rho -
i \eta ) \cr
- \lambda & 1 - \lambda^2 /2 & A \lambda^2 \cr
A \lambda^3 ( 1 - \rho - i \eta ) & - A \lambda^2 & 1 \cr } \right ]~~~~~ .
\end{equation}
Note the phases in the elements $V_{ub}$ and $V_{td}$.  These phases allow the
standard $V - A$ interaction to generate CP violation as a higher-order weak
effect.

The parameter $\lambda$ is measured by a comparison of strange particle
decays with muon decay and nuclear beta decay, leading to $\lambda \approx
\sin \theta \approx 0.22$, where $\theta$ is just the Cabibbo \cite{cab} angle.
The dominant decays of $b$-flavored hadrons occur via the element
$V_{cb} = A \lambda^2$.  The lifetimes of these hadrons and their semileptonic
branching ratios then lead to estimates in the range $A = 0.7 - 0.9$.
The decays of $b$-flavored hadrons to charmless final states allow one to
measure the magnitude of the element $V_{ub}$ and thus to conclude that
$\sqrt{\rho^2 + \eta^2} = 0.2 - 0.5$. The least certain quantity is the phase
of $V_{ub}$:  Arg $(V_{ub}^*) = \arctan(\eta/\rho)$.  We shall mention ways in
which information on this quantity may be improved, in part by indirect
information associated with contributions of higher-order diagrams involving
the top quark.

The unitarity of V and the fact that $V_{ud}$ and $V_{tb}$ are very close to
1 allow us to write $V_{ub}^* + V_{td} \simeq A \lambda^3$, or, dividing by a
common factor of $A \lambda^3$, $\rho + i \eta ~~ + ~~ (1 - \rho - i \eta)
= 1$. The point $(\rho,\eta)$ thus describes in the complex plane one vertex of
a triangle whose other two vertices are $(0,0)$ and $(0,1)$.  This triangle and
conventional definitions of its angles are depicted in Fig.~1.

\begin{figure}
\vspace{1.5in}
\caption{The unitarity triangle.  (a) Relation obeyed by CKM elements; (b)
relation obeyed by (CKM elements)/$A \lambda^3$}
\end{figure}

\subsubsection{Indirect information}

Indirect information on the CKM matrix comes from $B^0 - \bar B^0$ mixing and
CP-violating $K^0 - \bar K^0$ mixing, through the contributions of box diagrams
involving two charged $W$ bosons and two quarks of charge 2/3 $(u,~c,~t)$ on
the intermediate lines.  Evidence for the top quark with a mass of $m_t = 174
\pm 10~^{+13}_{-12}$ GeV/$c^2$ has recently been reported \cite{CDFtop},
reducing the errors associated with these box diagrams.

The original evidence for $B^0 - \bar B^0$ mixing came from the presence of
``wrong-sign'' leptons in $B$ meson semileptonic decays \cite{bbmix}.  The
splitting $\Delta m_B$ between mass eigenstates is proportional to $f_B^2 m_t^2
|V_{td}|^2$ times a slowly varying function of $m_t$.  Here $f_B$ is the $B$
meson decay constant.  The contributions of lighter quarks in the box diagrams,
while necessary to cut off the high-energy behavior of the loop integrals, are
numerically insignificant.

The CKM element $|V_{td}|$ is proportional to $|1 - \rho - i \eta|$.  Thus,
exact knowledge of $\Delta m_B,~f_B$ and $m_t$ would specify a circular arc
in the $(\rho,\eta)$ plane with center (1,0).  Errors on all these quantities
spread this arc out into a band. Present averages \cite{mixavg} give $(\Delta
m_B/\Gamma_B) = 0.71 \pm 0.07$. This value (close to 1) is nearly optimal for
observing CP-violating asymmetries in $B^0$ decays.

Similar box diagrams govern the parameter $\epsilon$ in CP-violating $K^0 -
\bar K^0$ mixing.  Here the dominant contribution to the imaginary part of the
off-diagonal mass matrix element is proportional to $f_K^2 m_t^2$ Im
$(V_{td}^2)$ times a slowly varying function of $m_t$.  Charmed quarks also
provide a small contribution.

The kaon decay constant is known: $f_K = 160$ MeV.  The imaginary part of
$V_{td}$ is proportional to $\eta(1-\rho)$.  Knowledge of $\epsilon$ thus
specifies a hyperbola in the $(\rho,\eta)$ plane with focus at $(1,0)$, which
is spread out into a band because of uncertainties in hadronic matrix elements.

\subsubsection{Allowed $(\rho,\eta)$ region}

\begin{figure}
\vspace{3.4in}
\caption{Contours of 68\% (inner curve) and 90\% (outer curve) confidence
levels for regions in the $(\rho,\eta)$ plane.  Dotted semicircles denote
central value and $\pm 1 \sigma$ limits implied by $|V_{ub}/V_{cb}| = 0.08 \pm
0.03$.  Plotted point corresponds to minimum $\chi^2 = 0.17$, while (dashed,
solid) curves correspond to $\Delta \chi^2 = (2.3,~4.6)$}
\end{figure}

Information on $|V_{ub}/V_{cb}|$ specifies a circular band in the $(\rho,\eta)$
plane.  When this constraint is added to those mentioned above, one obtains the
potato-shaped region shown in Fig.~2.  Here we have taken $m_t = 174 \pm 17$
GeV/$c^2$, $f_B = 180 \pm 30$ MeV, $(\rho^2 + \eta^2 )^{1/2} = 0.36 \pm 0.14$
(corresponding to $|V_{ub}/V_{cb}| = 0.08 \pm 0.03$), and $A = 0.79 \pm 0.09$
(corresponding to $V_{cb} = 0.038 \pm 0.005$). A parameter known as $B_K$
describes the degree to which the box diagrams dominate the $CP$-violating $K^0
- \bar K^0$ mixing.  We take $B_K = 0.8 \pm 0.2$, and set the corresponding
value for $B$ mesons equal to 1. A QCD correction \cite{BB} to the $B^0 - \bar
B^0$ mixing amplitude has been taken to be $\eta_{\rm QCD} = 0.6 \pm 0.1$.
Other parameters and fitting methods are as discussed in more extensive
treatments elsewhere \cite{HR,CKM}. Several parallel analyses \cite{BUR,AL}
reach qualitatively similar conclusions.

The best fit corresponds to $\rho \simeq 0,~\eta \simeq 0.36$, while at
90\% confidence level the allowed ranges are:
$$
\eta \simeq 0.3~~:~~~-0.4 \le \rho \le 0.4~~~;
$$
\begin{equation}
\rho \simeq 0~~:~~~\eta \simeq 0.3 \times 2^{\pm 1}~~~.
\end{equation}

A broad range of parameters gives an acceptable description of CP violation
in the kaon system.  The study of CP violation in $B$ decays could confirm
or disprove this picture.

\subsection{Modes of studying CP violation in $B$ decays}

Any manifestation of CP violation requires some sort of interference.  We
give two of the main examples under consideration for $B$ decays.  We then
discuss how charmed particles can provide useful information in both cases.

\subsubsection{Self-tagging decays}

Inequality of the rates for a process and its charge conjugate, such as $B^+
\to \pi^0 K^+$ and $B^- \to \pi^0 K^-$, would signify CP violation.  Under
charge conjugation, the weak phases change sign while the strong phases do not.
A rate difference can arise if both strong and weak phases are different in
two channels (here, $I = 1/2$ and $I = 3/2$).  Interpretation requires knowing
the strong phase shift difference $\delta \equiv \delta_{3/2} - \delta_{1/2}$.

\subsubsection{Decays to CP eigenstates}

Interference between a decay amplitude and a mixing amplitude can lead to rate
differences between decays of $B^0$'s and $\bar B^0$'s to CP eigenstates such
as $J/\psi K_S$ or $\pi^+ \pi^-$.  Here, no strong phase shift is needed to
generate an observable effect, and decay rate asymmetries can directly probe
angles of the unitarity triangle.  However, it is necessary to know the flavor
of the initial neutral $B$ meson.

\subsection{Final-state phases}

Several examples involving charmed particles can be instructive in how one
obtains final-state phase shift information from decay rates.  These
examples turn out to have parallels in the case of $B$ mesons, but the
cases of real interest for CP violation in the $B$ system turn out to be
somewhat more complex.

The decays $D \to \bar K \pi$ are characterized by the quark subprocess
$c \to s u \bar d$, which has $\Delta I = \Delta I_3 = 1$, and so there are two
final-state amplitudes, one with $I = 1/2$ and one with $I = 3/2$. The
amplitudes for decays to specific charge states can be written in terms of
isospin amplitudes as $A(D^+ \to \bar K^0 \pi^+) = A_{3/2}$; $A(D^0 \to K^-
\pi^+) = (2/3)A_{1/2} + (1/3)A_{3/2}$; $A(D^0 \to \bar K^0 \pi^0) = \sqrt{2}
(A_{3/2} - A_{1/2})/3$.  The amplitudes then obey a triangle relation, and by
considering the observed rates one finds the relative phase of the $I = 1/2$
and $I = 3/2$ amplitudes to be around 90$^{\circ}$ \cite{stonec}.  This is
likely to indicate the importance of resonant structure. The $I = 1/2$ channel
is ``non-exotic'' (it can be formed of a quark-antiquark state), while the $I =
3/2$ channel is ``exotic,'' requiring at least two quarks and two antiquarks.
No resonances have been seen in exotic channels, while there is an $I = 1/2~K
\pi$ resonance just around the mass of the $D$ meson \cite{kpires}.

Triangle constructions similar to that mentioned above indicate that the
relative phase of $I = 1/2$ and $I = 3/2$ amplitudes in $D \to \bar K^* \pi$
appears to be about 90$^{\circ}$, while it appears to be about 0 in $D \to \bar
K \rho$.  This difference may be due to details of resonance couplings, but
could not have been anticipated {\it a priori}.  It illustrates the importance
of actual measurements rather than theoretical prejudices in the evaluation
of final-state phase shift differences.

The decays $D \to \pi \pi$ are governed by the subprocess $c \to d u \bar d$
(or $c \to u$ penguin subprocesses).  The $\Delta I = 1/2$ transitions lead
to an $I = 0~\pi \pi$ final state, while the $\Delta I = 3/2$ transitions lead
to an $I = 2~\pi \pi$ final state.  Again, a triangle relation holds between
amplitudes, and the $I = 0$ and $I = 2$ amplitudes are found \cite{Selen} to
have a relative phase consistent with 90$^{\circ}$.

The decays $B \to \bar D \pi$ involve the quark subprocess $\bar b \to \bar c
u \bar d$ and so their isospin analysis parallels that of $D \to \bar K \pi$.
It has recently been concluded \cite{Hitoshi} that present data are consistent
with a relative phase shift of zero between the $I = 1/2$ and $I = 3/2$
amplitudes.

The decays $B \to K \pi$ involve the quark subprocesses $\bar b \to \bar s
u \bar u$ and $\bar b \to \bar s$ (penguin processes), and thus are
characterized by both $\Delta I = 0$ and $\Delta I = 1$ transitions.  The
$\Delta I = 0$ transitions can lead only to an $I = 1/2$ final state,
while the $\Delta I = 1$ transitions lead to both $I = 1/2$ and $I = 3/2$
final states.  Four $B \to K \pi$ decay amplitudes then can be expressed in
terms of two $I = 1/2$ and one $I = 3/2$ reduced amplitude, leading to a
quadrangle relation \cite{Quad}. Suggestions have been made \cite{BPP}
for incorporating information from $B \to \pi \pi$ decays with the help
of flavor SU(3) and untangling various final-state phases in the $K \pi$
channel.

\subsection{Flavor tagging in neutral $B$ decays}

As mentioned above, the decays of neutral $B$ mesons to CP eigenstates can
provide crisp information on angles in the unitarity triangle if one can
``tag'' the flavor of the decaying $B$ at the time of its production.  One
method for doing this \cite{GNR} relies on the correlation of a neutral $B$
with a charged pion.

This method \cite{SN} is already in use for tagging neutral $D$ decays.  The
charged $D^*$ resonance is far enough above the neutral $D$ that the decays
$D^{*+} \to \pi^+ D^0$ and $D^{*-} \to \pi^- D^0$ are kinematically allowed.
Here one is interested in whether a given final state has arisen from
mixing or from the doubly-suppressed process $c \to d u \bar s$.

In the case of $B$ mesons, the $B^*$ is only 46 MeV above the $B$, so the
decay $B^* \to B \pi$ is kinematically forbidden.  Nonetheless, one can
expect non-trivial correlations between the flavor of a produced $B$ and
a pion nearby in phase space, either as a result of correlations in the
fragmentation process or through the decays of resonances above the $B^*$.
In both cases, the corresponding physics for charmed particles is easy to
study and will provide interesting information.

\subsection{Pion -- $B$ correlations}

The pion-$B$ correlation in a fragmentation picture is illustrated in Fig.~3.
When incorporated into a neutral $B$ meson, a $\bar b$ quark is ``dressed''
with a $d$, leading to a $B^0$.  The next quark down the rapidity chain is a
$\bar d$, which will appear in a pion of positive charge.  Similarly, a $\bar
B^0$ is more likely to be correlated with a $\pi^-$.

\begin{figure}
\vspace{1.5in}
\caption{Quark graphs illustrating pion-$B$ correlations.  Fragmentation
of a $\bar b$ quark leads to a $B^0$ and a nearby $\pi^+$, while fragmentation
of a $b$ quark leads to a $\bar B^0$ and a nearby $\pi^-$.}
\end{figure}

The existence of this correlation in CDF data is still a matter of some
debate.  It would be interesting to see if it exists for charmed particles.
One would have to subtract out the contribution of $D^*$ decays, of course.

\subsection{$B^{**}$ resonances and their charmed equivalents}

A $B^0$ or $B^{*0}$ can resonate with a positive pion, while a $\bar B^0$ or
$\bar B^{*0}$ can resonate with a negative pion. The combinations $B^0 \pi^-$
and $\bar B^0 \pi^+$ are exotic, and not expected to be resonant.

The lowest-lying resonances which can decay to $B \pi$ or $B^* \pi$ are
expected to be the P-wave $\bar b q$ states.  We call them $B^{**}$ (to
distinguish them from the $B^*$'s).  The expectations for masses of these
states \cite{GNR,EHQ}, based on extrapolation from the known $D^{**}$
resonances, are summarized in Table 1.

\begin{table}
\begin{center}
\caption{P-wave resonances of a $b$ quark and a light ($\bar u$
or $\bar d$) antiquark}
\medskip
\begin{tabular}{c c c} \hline
$J^P$  &    Mass     &  Allowed final \\
       & (GeV/$c^2$) &     state(s)   \\ \hline
$2^+$  & $\sim 5.77$ &  $B \pi,~B^* \pi$ \\
$1^+$  & $\sim 5.77$ &     $B^* \pi$  \\
$1^+$  & $ < 5.77$   &     $B^* \pi$  \\
$0^+$  & $ < 5.77$   &      $B \pi$   \\ \hline
\end{tabular}
\end{center}
\end{table}

The known $D^{**}$ resonances are a $2^+$ state around 2460 MeV/$c^2$, decaying
to $D \pi$ and $D^* \pi$, and a $1^+$ state around 2420 MeV/$c^2$, decaying to
$D^* \pi$.  These states are relatively narrow, probably because they decay
via a D-wave.  In addition, there are expected to be much broader (and probably
lower) $D^{**}$ resonances:  a $1^+$ state decaying to $D^* \pi$ and a $0^+$
state decaying to $D \pi$, both via S-waves.

Once the masses of $D^{**}$ resonances are known, one can estimate those of the
corresponding $B^*$ states by adding about 3.32 GeV (the quark mass difference
minus a small binding correction).  Adding a strange quark adds about 0.1 GeV
to the mass.  Partial decay widths of $D^{**}$ states are also related to those
of the $B^{**}$'s \cite{EHQ}.  Thus, the study of excited charmed states can
play a crucial role in determining the feasibility of methods for identifying
the flavor of neutral $B$ mesons.

\section{Strange $B$'s}

\subsection{Production}

It is important to know the ratios of production of different $B$ hadrons:
$B^+~:~B^0~:~B_s~:~\Lambda_b$.  These ratios affect signals for mixing and
the dilution of flavor-tagging methods.  Aside from effects peculiar to the
decays $D^* \to D \pi$, one should have similar physics in the ratios
$D^+~:~D^0~:~D_s~:~\Lambda_c$.

\subsection{Masses}

It appears that the $B_s$ states are about 90 MeV above the $B$'s \cite{EHQ}.
One predicts a similar splitting for the strange and nonstrange vector mesons
\cite{RW}. The corresponding splittings for charmed particles are about 100 MeV
for both pseudoscalar and vector mesons, as well as for the observed P-wave
levels.  This leads to a more general question:  How much mass does a strange
quark add?  This is an interesting ``isotope effect'' which in principle could
probe binding effects in the interquark force.

\subsection{$B_s - \bar B_s$ mixing}

The box diagrams which lead to $K^0 - \bar K^0$ and $B^0 - \bar B^0$ mixing
also mix strange $B$ mesons with their antiparticles. One expects $(\Delta
m)|_{B_s}/(\Delta m)|_{B_d} = (f_{B_s}/f_{B_d})^2 (B_{B_s}/B_{B_d})
|V_{ts}/V_{td}|^2$, which should be a very large number (of order 20 or more).
Thus, strange $B$'s should undergo many particle-antiparticle oscillations
before decaying.

The main uncertainty in an estimate of $x_s \equiv (\Delta m/
\Gamma)_{B_s}$ is associated with $f_{B_s}$.  The CKM elements $V_{ts} \simeq
-0.04$ and $V_{tb} \simeq 1$ which govern the dominant (top quark) contribution
to the mixing are known fairly well. We show in Table 2 the dependence of
$x_s$ on $f_{B_s}$ and $m_t$. To measure $x_s$, one must study the
time-dependence of decays to specific final states and their charge-conjugates
with resolution much less than the $B_s$ lifetime (about 1.5 ps).

\begin{table}
\begin{center}
\caption{Dependence of mixing parameter $x_s$ on top quark mass and
$B_s$ decay constant.}
\medskip
\begin{tabular}{c c c c} \hline
\null \qquad $m_t$ (GeV/$c^2$)&  157  &  174  &  191  \\ \hline
$f_{B_s}$ (MeV)               &       &       &       \\
150                           &  7.6  &  8.9  & 10.2  \\
200                           & 13.5  & 15.8  & 18.2  \\
250                           & 21.1  & 24.7  & 28.4  \\ \hline
\end{tabular}
\end{center}
\end{table}

\section{Heavy meson decay constants}

\subsection{The $D_s$}

{\it Direct measurements} are available so far only for the $D_s$ decay
constant.  The WA75 collaboration \cite{WA75} has seen 6 -- 7 $D_s \to \mu \nu$
events, and Fermilab E653 and the BES detector at the Beijing Electron-Positron
Collider (BEPC) also have a handful.  The CLEO Collaboration \cite{FDSCLEO} has
a much larger statistical sample; the main errors arise from background
subtraction and overall normalization (which relies on the $D_s \to \phi \pi$
branching ratio).  The actual measurement is $r \equiv B(D_s \to \mu \nu)/B(D_s
\to \phi \pi) = 0.245 \pm 0.052 \pm 0.074$.

A better measurement of $B(\phi \pi) \equiv B(D_s \to \phi \pi)$ is sorely
needed. One method \cite{MS} is to apply factorization \cite{fact} to the decay
$B \to D_s D$, where $D_s \to \phi \pi$, to obtain the combination $f_{D_s}^2
B(\phi \pi)$.  Since $r \propto f_{D_s}^2 / B(\phi \pi)$, one can extract both
the decay constant and the desired branching ratio.  Using this and other
methods, Muheim and Stone \cite{MS} estimate $f_{D_s} = 315 \pm 45$ MeV and
$B(\phi \pi) = (3.6 \pm 0.6)\%$.

The large value of $f_{D_s}$ implies a branching ratio of about 9\% for
$D_s \to \tau \nu_\tau$.  This is good news for experiments \cite{P872}
contemplating the production of $\nu_\tau$ in beam dumps.

\subsection{The charged $D$}

By searching for the decay $D \to \mu \nu$ in the decays of $D$ mesons produced
in the reaction $e^+ e^- \to \psi(3770) \to D^+ D^-$, the Mark III
collaboration has obtained the upper limit \cite{MKIII} $f_D < 290$ MeV (90\%
c.l.).  The BES detector at Beijing should be able to improve upon this limit,
which is not far above theoretical expectations \cite{PASCOS,BLS,LAT}.

The CLEO measurement of $f_{D_s}$ mentioned above relied on photon-$D_s$
correlations in the decay $D_s^* \to D_s \gamma$.  One may be able to search
for the decay $D^+ \to \mu \nu$ by looking for the $\pi^0 - D^+$ correlation
in the decay $D^{*+} \to D^+ \pi^0$ \cite{FD}.

\subsection{$B$ Meson decay constants}

If $f_B$ were better known, the indeterminacy in the $(\rho,\eta)$ plane
associated with fits to CKM parameters would be reduced considerably.  We show
in Fig.~4 the variation in $\chi^2$ for the fit described in Sec.~3.1 when
$f_B$ is taken to have a fixed value.  An acceptable fit is obtained for a wide
range of values, with $\chi^2 = 0$ for $f_B = 153$ and 187 MeV.

\begin{figure}
\vspace{4in}
\caption{Variation of $\chi^2$ in a fit to CKM parameters as a function
of $f_B$.}
\end{figure}

The reason for the flat behavior of $\chi^2$ with $f_B$ is illustrated in
Fig.~5.  The dashed line, labeled by values of $f_B$, depicts the $(\rho,\eta)$
value for the solution with minimum $\chi^2$ at each $f_B$.  The product
$|1 - \rho - i \eta| f_B$ is constrained to be a constant by $B^0 - \bar B^0$
mixing.  The product $\eta (1 - \rho)$ is constrained to be constant by
the value of $\epsilon$.  The locus of solutions to these two conditions
lies approximately tangent to the circular arc associated with the
constraint on $|V_{ub}/V_{cb}|$ for a wide range of values of $f_B$.

\begin{figure}
\vspace{3in}
\caption{Locus of points in $(\rho,\eta)$ corresponding to minimum $\chi^2$ for
fixed values of $f_B$.  Circular arcs depict central value and $\pm 1 \sigma$
errors for $|V_{ub}/V_{cb}|$.  Solid dots denote points with $\chi^2 = 0$.}
\end{figure}

The uncertainty in $f_B$ thus becomes a major source of uncertainty in $\rho$,
which will not improve much with better information on $|V_{ub}/V_{cb}|$.
Fortunately, several estimates of $f_B$ are available, and their reliability
should improve.

{\it Lattice gauge theories} have attempted to evaluate decay constants
for $D$ and $B$ mesons.  A representative set \cite{BLS} is
$$
f_B = 187 \pm 10 \pm 34 \pm 15~~{\rm MeV}~~~,
$$
$$
f_{B_s} = 207 \pm 9 \pm 34 \pm 22~~{\rm MeV}~~~,
$$
$$
f_D = 208 \pm 9 \pm 35 \pm 12~~{\rm MeV}~~~,
$$
\begin{equation}
f_{D_s} = 230 \pm 7 \pm 30 \pm 18~~{\rm MeV}~~~,
\end{equation}
where the first errors are statistical, the second are associated with fitting
and lattice constant, and the third arise from scaling from the static $(m_Q =
\infty)$ limit.  The spread between these and some other lattice
estimates \cite{LAT} is larger than the errors quoted above, however.

{\it Quark models} can provide estimates of decay constants and their ratios.
In a non-relativstic model \cite{ES}, the decay constant $f_M$ of a heavy meson
$M = Q \bar q$ with mass $M_M$ is related to the square of the $Q \bar q$ wave
function at the origin by $f_M^2 = 12 |\Psi(0)|^2 / M_M$.  The ratios of
squares of wave functions can be estimated from strong hyperfine splittings
between vector and pseudoscalar states, $\Delta M_{\rm hfs} \propto
|\Psi(0)|^2/m_Q m_q$.  The equality of the $D_s^* - D_s$ and $D^* - D$
splittings then suggests that
\begin{equation}
f_D/f_{D_s} \simeq (m_d/m_s)^{1/2} \simeq 0.8 \simeq f_B/f_{B_s}~~~,
\end{equation}
where we have assumed that similar dynamics govern the light quarks bound to
charmed and $b$ quarks.  In lattice estimates these ratios range between
0.8 and 0.9.

An improved measurement of $f_{D_s}$ and a first measurement of $f_D$ could
provide a valuable check on predictions of various theories and could help
pin down $B$ meson decay constants, since ratios are expected to be more
reliably predicted than individual constants \cite{BG}.

\section{Charmed baryon spectra}

The $\Lambda_c$ baryon is a particularly simple object in heavy-quark symmetry,
since its light-quark system consists of a $u$ and $d$ quark bound to a state
$[ud]$ of zero spin, zero isospin, and color antitriplet.  Comparisons with the
$\Lambda_b = b[ud]$ and even with the $\Lambda = s[ud]$ are thus particularly
easy.

The $[ud]$ diquark in the $\Lambda$ can be orbitally excited with respect to
the strange quark.  The $L = 1$ excitations consist of a fine-structure
doublet, the $\Lambda(1405)$ with spin-parity $J^P = 1/2^-$ and the
$\Lambda(1520)$ with $J^P = 3/2^-$.  The spin-weighted average of this
doublet is 366 MeV above the $\Lambda$.  These states are illustrated on the
left-hand side of Fig.~6.

\begin{figure}
\vspace{4.25in}
\caption{Ground states and first orbital excitations of $\Lambda$ and
$\Lambda_c$ levels.}
\end{figure}

Within the past couple of years candidates have been observed \cite{EXLC} for
a corresponding $L = 1$ doublet of charmed baryons.  These are illustrated on
the right-hand side of Fig.~6.  The lower-lying candidate, 308 MeV above the
$\Lambda_c$, decays to $\Sigma_c \pi$, while the higher-lying candidate, 342
MeV above the $\Lambda_c$, does not appear to decay to $\Sigma_c \pi$, but
rather to $\Lambda_c \pi \pi$.  This pattern can be understood \cite{Pes}
if the lower candidate has $J^P = 1/2^-$ and the higher has $J^P = 3/2^+$.
The lower state can decay to $\Sigma_c \pi$ via an S-wave, while the higher
one would have to decay to $\Sigma_c \pi$ via a D-wave.  It would have no
trouble decaying to $\Sigma_c^* \pi$ via an S-wave, however.  The predicted
$\Sigma_c^*$, with $J^P = 3/2^+$, has not yet been identified.

The spin-weighted average of the excited $\Lambda_c$ states is 331 MeV above
the $\Lambda_c$, a slightly smaller excitation energy than that in the
$\Lambda$ system.  The difference is easily understood in terms of reduced-mass
effects.  The ${\bf L \cdot S}$ splittings appear to scale with the inverse of
the heavy quark ($s$ or $c$) mass.

The corresponding excited $\Lambda_b$ states probably lie 300 to 330 MeV
above the $\Lambda_b(5630)$, with an ${\bf L \cdot S}$ splitting of about 10
MeV.

\section{Lifetime differences}

Charmed particle lifetimes range over a factor of ten, with
\begin{equation}
\tau(\Xi_c^0) < \tau(\Lambda_c) < \tau(\Xi_c) \simeq \tau(D^0) \simeq
\tau(D_s) < \tau(D^+)~~~.
\end{equation}

Effects which contribute to these differences \cite{lifes} include (a) an
overall nonleptonic enhancement from QCD \cite{enh}, (b) interference when at
least two quarks in the final state are the same \cite{int}, (c) exchange and
annihilation graphs, e.g. in $\Lambda_c$ and $\Xi_c^0$ decays \cite{exch}, and
(d) final-state interactions \cite{fsi}.

In the case of $B$ hadrons, theorists estimate that all these effects shrink
in importance to less than ten percent \cite{blife}.  However, since the
measured semileptonic branching ratio for $B$ decays of about 10 or 11\%
differs from theoretical calculations of 13\% by some 20\%, one could easily
expect such differences among different $b$-flavored hadrons.  These could
arise, for example, from final-state interaction effects.  As mentioned
earlier \cite{BPP}, there are many tests for such effects possible in the
study of decays of $B$ mesons to pairs of pseudoscalars.

\section{Summary}

Charmed particles are a rich source of information about what to expect in the
physics of particles containing $b$ quarks, in addition to being interesting in
their own right.

Some properties of charmed particles are expected to be very close to those of
$B$ hadrons, such as excitation energies.  Others are magnified in the case of
charm, being proportional to some inverse power of the heavy quark mass.

Charmed particles are easier to produce than $B$ hadrons in a hadronic
environment (and in photoproduction), and so are a natural area of study for
fixed-target experiments such as those being performed and planned at Fermilab.
The high-statistics study of charmed particles could have a broad impact on
fundamental questions in particle physics.

\end{document}